# Mapping an audience centric World Wide Web: A departure from hyperlink analysis

Harsh Taneja
Missouri School of Journalism, University of Missouri

**Abstract**

This paper argues that maps of the Web's structure based solely on technical infrastructure such as hyperlinks may bear little resemblance to maps based on Web usage, as cultural factors drive the latter to a larger extent. To test this thesis, the study constructs two network maps of 1000 globally most popular Web Domains, one based on hyperlinks and the other using an "audience centric" approach with ties based on shared audience traffic between these domains. Analyses of the two networks reveal that unlike the centralized structure of the hyperlinks network with few dominant "core" websites, the audience network is more decentralized and clustered to a larger extent along geo-linguistic lines

**Harsh Taneja** (PhD, Northwestern University) is an Assistant Professor in the School of Journalism at University of Missouri. His research focuses on media audiences, and explores connections between patterns of media use and the accompanying institutional and social structures.
**Address:** 181-C Gannett Hall, Missouri School of Journalism, Columbia, MO, USA −65211-1200.
**Email:** harsh.taneja@gmail.com

# Mapping an audience centric World Wide Web: A departure from hyperlink analysis

The Internet and specifically the World Wide Web (henceforth, Web) has emerged as a platform with a massive capacity for global information exchange with social, economic, political and cultural consequences. Hence, scholars in disciplines ranging from physics and information sciences to geography and communication studies are interested in mapping its structure. All such maps attempt at providing a factual reality of the Internet and include "architectural plans, engineering blueprints, anatomical drawings, and statistical graphics" (Dodge, 2008: 352).

A common approach in the social sciences to mapping the Internet is to create networks of the structure of the Web based on hyperlinks between Websites. Hyperlinks reflect the behavior of webmasters and any imagery of the Internet based on technical infrastructure alone is insufficient to map the heterogeneous contours of actual global Web usage, especially as Internet access has both broadened and deepened globally. Therefore, maps of the Web, based solely on hyperlinks, provide a partial representation of the medium's structure. Recognizing this limitation, recent studies have mapped the Web using audience centric approaches, either based on shared audiences between websites (e.g., Taneja and Webster, 2016, Taneja and Wu, 2014) or user clickstreams (e.g., Wu and Ackland, 2015). This paper empirically compares audience centric structure of the Web with its' hyperlink structure.

Specifically, this study maps the Web using two simultaneously obtained datasets on the world's 1000 most popular websites, one an "audience map", based on shared audience traffic between these websites and the other on hyperlinks between them. A network analysis of both these maps suggests that cultural factors, such as linguistic and geographic similarity between websites, explain the audience map to a larger extent than they explain the "hyperlinks map". These findings have implications for popular and scholarly imaginations of the global Internet.

## The global Internet structure

### Hyperlink analysis

Maps of hyperlinks have contributed to popular and scholarly imaginations of the global Internet for about two decades. Based on counting hyperlinks between pages, these maps conceptualize the Web as a network with Web pages as nodes, where any two pages are tied to one another if they have hyperlinks between them. These nodes can be individual Web pages, or their higher-level aggregations such as Websites or "top level domains" (e.g.".com", ".de"). The earliest such studies, found that the Web has a centralized "bow-tie" structure with a strongly connected "core" component and several peripheral weakly connected components (e.g., Barabasi and Albert, 1999 Broder et al., 2000;). As the Web has grown enormously in size, recent studies also find hyperlink networks to exhibit similar properties irrespective of whether one analyzes individual Web pages (Meusel et al., 2014) or aggregations of pages such as Websites, or "pay level domains" (the latter refers to sub domains of top level domains that webmasters generally pay for, see Lehmberg et al., 2014). Hyperlink analysis at this scale has consistently suggested that the Web has a "core/ periphery" structure. In other words, a few (core) sites receive links from most other sites (periphery) resulting in the Web as a highly centralized network. The World Systems Theory (WST) is often used to explain the observed core-periphery structure.

Classically, WST has been applied to explain the dominance of certain nations over others at different times in world history, and it divides most countries of the world into core, periphery, and semi-periphery with the countries in the core dominating the others due to their superior economic prowess (Chase-Dunn and Grimes, 1995). Based on this classification, WST explains the dependency of third world countries over their first world counterparts, and their inequality and imbalance in a variety of areas. Originally developed as a theory primarily to explain flows of goods and people between countries, scholars have also used WST to account for the highly centralized structure of global information flows, as evidenced in Web hyperlink networks (Barnett and Park, 2005).

Using data collected in 1998, Barnett et al. (2001) constructed a country-to-country hyperlink network aggregated on the basis of the sampled sites' top-level domains. For instance, they assigned sites ending with a ".co.uk" to United Kingdom and sites with ".de" extension to Germany. Through network analysis of 29 OECD countries' domains, their study concluded that US was the most central node and countries with higher GDP constituted the "core" of the country to country hyperlinks network with low GDP countries at the periphery. Expanding this inquiry to 47 countries using data from 2003 and also measuring bilateral bandwidth capacity (of physical network components such as cables and fibers) in addition to hyperlinks, Barnett and Park (2005) again found the Web to exhibit a core-periphery structure.

> The structure of both networks is consistent with world system theory… Both networks can be arrayed along a single core-periphery dimension with the U.S. and the wealthier nations of Western Europe at the center and the poor less developed nations of Latin America, Asia and Africa along the margins. (Barnett and Park, 2005: 1123)

Replicating this analysis on hyperlink data obtained in 2009, Park et al. (2011) found the overall network to be structurally quite similar to the one in 2003 that exhibited the core-periphery hierarchies posited by the WST. They found that the whole network overall had become more centralized in favor of developed countries such as the G7 and the US. In a related study, Barnett et al.,(2011) used an expanded sample and improved methodology (unlike Park et al., 2011, they were able to include and assign generic domains such as ".com" to their countries of origin) and found the centrality of US to have further increased 30 times; this finding validated their earlier studies that interpreted Web hyperlinks to have a core-periphery structure. Such studies posited that a country's potential for development and its dependency on or domination of other countries predicted its position on the global Internet structure.

Studies focusing on specific genres of Websites have also found similar trends. Analyzing hyperlinks provided by 223 news Websites in 70 countries worldwide on international news stories, a study found that Websites in peripheral countries mostly linked to Websites in the developed countries and the latter hardly provided any links to the former (Himelboim, 2010). Similarly, a recent study of hyperlinks between the Websites of the 1000 most prominent universities globally revealed that universities from United States, Germany and United Kingdom occupy central positions (Barnett et al., 2014).

Studies in this research tradition find cultural factors to be "weak second predictors" of global hyperlinks structure, compared to economic factors. For instance, Barnett and Sung (2005) find the GDP of a country to be a much better predictor of its centrality on the Internet than any dimension of culture and conclude that "[a]s suggested by World Systems Theory, the economy rather than culture is the primary determinant of the structure of international hyperlink flows" (Barnett and Sung, 2005: 230).

A related explanation, often advanced for the observed core-periphery structure of hyperlinks, especially the centrality of US and Western Europe (including the UK) on the Web, is the central place that the English language occupies in the modern world. Studies find that blogs generally link to same-language blogs and to English language media (Hale, 2012). Extending this to the wider Web, a large scale Web crawl by Google on 2008 data demonstrated that most Websites almost exclusively link to other Websites in the same language, with one exception; they also link to pages in English (Daniel and Josh, 2011). The Google study further found that for some languages, a rather high proportion of pages linked to in English. For instance, 43% of all hyperlinks from Web pages in Urdu linked to pages in English. Historically more online content has been created in English than in other languages (Hecht and Gergle, 2010; Herring et al., 2007).

In recent years, with Internet penetration in the global South both deepening and widening, the percentage of native English speakers on the Web has fallen dramatically and English language content occupies a much smaller proportion of all available Webpages. One study estimated that between 1996 and 2006, the proportion of native English speakers online reduced from 80% to fewer than 30% and the percentage of English language Web pages steadily declined from 75% to 45% (cited in Zuckerman, 2013). Since 2006, with Internet penetration growing in China, Brazil and India, these numbers would have come down even further. However, studies such as the crawl by Google find English to be highly central on the global language hyperlink network, even after controlling for the number of pages available in each language (Daniel and Josh, 2011). More recent studies also find that despite the growth in pages in other languages, English continues to be highly central in online and offline global language networks (Ronen et al., 2014).

In summary, hyperlinks networks of the Web consistently reveal that a few western countries (especially the US) occupy highly central positions on the Web, suggesting unequal and asymmetric global flows of digital information. The structure of the Web as inferred exclusively from network analysis of hyperlinks, does not reflect how actual usage shapes structural relationship between Websites. I expect this divergence between usage and hyperlinks based on two bodies of literature; studies on why Webmasters hyperlink, and studies on global cultural consumption. In the sections that follow, I review each of these in turn and also note an emergent methodological complexity with crawling the Web for hyperlinks.

**Motivations for hyperlinking**

Webmasters provide hyperlinks for several purposes. Other than influencing user traffic, Webmasters utilize hyperlinks to reinforce offline relations, to signal legitimacy, mobilize resources or at times to even express power and legitimacy (De Maeyer, 2013). As an example, bloggers, especially writers of political blogs, tend to link to other blogs and pages that are on the same side of the political spectrum (Adamic and Glance, 2004). This behavior is largely attributed to either the intrinsic nature of bloggers to favor their own kind or due to network forces that reflect the "relative popularity of one ideology over the other within the population from which bloggers are drawn" (Ackland and Shorish, 2009: 397).

It is common to find hyperlinks connecting sites owned by the same company even though they are in different languages. For instance, CNN provides a large number of hyperlinks between its Spanish language Mexican edition and its English language US and International Editions. Likewise, most Wikipedia pages have hyperlinks to most other Wikipedia pages that cover the concept in other languages. However, it is unlikely that a Spanish-speaking user from Spain who accesses information about New York City in Spanish would click on the link that provides information on New York City in other languages, unless she finds the information in Spanish inadequate and is comfortable with comprehending another language. Such situations indicate instances where links are most likely provided to signal affiliation and the link providers may not expect their website users to actually click through them.

In counting links alone, researchers have very limited information about the inherent motivations of the link provider. Moreover, two site owners who own the same type of site may have different motivations for providing hyperlinks. The simplest dichotomy being that a link could signal either praise or criticism (De Maeyer, 2012). Further, either the link itself or the linked resource may reflect the intention of the link provider. Finally, webmasters may employ different linking strategies for Websites based on their genres. A news Website may link to an influential Website to signal legitimacy whereas a personal homepage may have links representing pages and organizations personally relevant to the author. Despite these limitations, scholars continue to map the Web as network of hyperlinks since "they are among the best data available as they can be observed passively, are publicly available and possess a similarity to citation" (Hale, 2012: 136). Summarizing the need to augment hyperlink analysis with other methods, a study notes,

> ...several methods need to be employed to examine the reasons developers of Websites form a network with other sites via hyperlinks: survey, in-depth interviews, observation, comparative analysis of Website contents and other network data would contribute to an understanding of the social relationships among the network's components. (Park, 2003: 58)

Given the different motivations for providing hyperlinks, it is likely that hyperlinks based maps differ from actual Web usage. The literature on global cultural consumption, reviewed next, indeed suggests that usage is driven by cultural factors.

**Global Web use and culturally defined markets**

Early scholarship propounded a vision of global cultural consumption consistent with the core-periphery structure observed in maps of the Web based on hyperlinks. Scholars raised concerns that unrestricted global flow of information is a harbinger of "cultural imperialism" and feared that cultural products from large and wealthy western nations swamp local offerings and homogenize cultural consumption (Schiller, 1969). Others associate media imperialism with the growing clout of major transnational media corporations aided by deregulation (Arsenault and Castells, 2008; Herman and McChesney, 2001). Economists use market forces to explain these imbalanced one way flows, positing that larger nations have a competitive advantage over smaller ones in the international trade of cultural products (Wildman, 1994).

Most students of global culture expect something quite different. They believe in the power of "cultural proximity," and the continued ability of cultural structures like language and geography to shape audiences because people prefer content that is closer to their own culture. Given a choice, audiences prefer products that are closer to their culture than those from abroad (Pool, 1977). Hence, when seen in the aggregate, global media consumption manifests as an aggregation of many "culturally defined markets", aligned on geo-linguistic lines (Straubhaar, 1991, 2007). Although the theory of cultural proximity developed to explain global consumption of television and films, language and geography explain global Web use patterns to a large extent (e.g., Taneja and Wu, 2014). In fact, these cultural factors predict Web use better than hyperlinks do (Taneja and Webster, 2016).

The World Wide Web has been designed and conceived of in popular and scholarly discourse, as an inherently global mass medium (Castells, 1996). Any user with access to an Internet enabled device, can potentially access the Web irrespective of location, with few exceptions. Remarkably, in the last decade, Web usage in the Global South has seen enormous growth. For instance, between 2005 and 2015, China's Internet user base increased from 112 million users to 725 million (Internet Live Stats). Despite its size, the Web is an almost fully connected network due to its hyperlinked architecture, which enables users to easily navigate to any corner of the Web. It has "the *small world* property, which [ensures] that any two nodes are likely to be connected… by a relatively short path of nodes—in the case of the Web, the path length is about 19" (Barabási, 2013, p. 2). Further the Web is also a *scale free network,* that is, the addition of more pages does not increase the average distance between any two pages in the same proportion, because newest sites tend to hyperlink to the few established popular sites that already have a large number of links (Barabási and Albert, 1999). These popular sites tend to be based in Western countries where the Web became popular first. Hence even as the Web has grown, the average degree of separation between pages has remained relatively small (Barabási, 2009).

Ease of navigation through hyperlinks and other tools such as search engines, recommendation systems and machine translation have motivated visions of a completely connected, globalized world of Web users. Despite these possibilities, audiences continue to pay "disproportionate attention to phenomena that unfold nearby and directly affect ourselves, our friends and our families" (Zuckerman,

2013:19). Twitter ties form between people in close geographic proximity (Takhteyev et al., 2012). Likewise, people are more likely to email someone who shares their culture than someone from another "civilization" (State et al., 2015). Each language edition of Wikipedia differs from others in the concepts they cover and even when they cover the same concept, their content can be quite different (Hecht and Gergle, 2010). Thus, despite ubiquitous access, both language and physical geography significantly shape Web usage.

Consistent with the cultural proximity thesis, studies find that global information and communication flows on global telecommunications networks (including the Internet) have decentralized as well as diversified over time. With increasing access to and use of information and communication technologies around the world, people in the erstwhile "peripheral" countries communicate with one another directly, and such communication is no longer routed through core countries. One study observed such changes in global telecommunications traffic between 1989 and 1999, when the overall network centralization decreased in 1999 and regional clusters emerged (Lee et al., 2007). Likewise, another study analyzed global telecommunication traffic networks between 1978 and 2009 and found that countries erstwhile at the periphery, such as China, India, Philippines and Guatemala and the former "Eastern Bloc", all have become more central over time (Barnett, 2012).

These studies suggest that the core-periphery structure of the Web, as suggested by hyperlinks analysis, is an inaccurate representation of how global usage patterns shape the Web. Further, the nature of the Web itself has changed somewhat, which impacts the ability of Web-crawlers to accurately capture all hyperlinks. Web 2.0 environments, especially social media platforms allow ordinary Web users to provide hyperlinks while sharing content, most of which is semi-private (I.e., visible only to a user's social network) and therefore Web crawlers cannot retrieve most such content. This methodological difficulty further limits the ability of crawled hyperlink graphs to accurately represent user behavior, especially given the ubiquity of sharing via social media Websites.

**Audience centric structures of the Web**

The possible divergence between hyperlinks and usage, has motivated "audience centric approaches" to map global web usage. However, so far only two recent studies have empirically tested for associations between hyperlinks and usage (Wu and Ackland, 2014, Barnett and Park, 2014); both find that usage and hyperlinks diverge. Both these studies rely on usage estimates from Alexa.com, a web information company, which estimates traffic from a self-selected sample of Web users and Website providers. Barnett and Park (2014) analyze usage at a national level, derived from commonalities in 100 most popular sites in each country based on Alexa's country wise Website traffic rankings. Wu and Ackland (2014) compared hyperlinks and outgoing- clickstream traffic between the same set of 980 websites, the major limitation being that, Alexa provides clickstream traffic for each site from/to a maximum of ten sites. Notably both these studies hint at culture being a greater driver of usage than it is for hyperlinks. The present study extends these studies with more extensively measured web usage data.

An audience centric approach, as used in this study, considers connections between Websites on the basis of their shared or "duplicated" audiences (Webster and Ksiazek, 2012) not just clickstreams. Audience duplication is the extent to which the same audiences consume two media outlets. For instance, on any given day, if out of 100 people in a population, 20 people watched both FOX News and CNN, the audience duplication between them would be 20% for that day. Likewise, duplication can be calculated for all possible pairs of media outlets for a given audience. This results in a symmetric audience duplication matrix, where the elements $A_{ij}$ represent the extent to which media outlets i and j have audiences in common. Such a matrix can be analyzed further to identify clusters of media outlets that have audiences in common (Ksiazek, 2011 describes this procedure in detail).

An early application of audience duplication was in identifying user defined program types, the subsets of television programs that were watched by the same set of users (Webster, 1985). Taneja and

Wu (2014) adapted audience duplication to study global web usage and their findings suggest that cultural factors drive such user centric maps of the Web. Unlike hyperlink analysis, which generally suggests the dominance of a small number of Websites (usually from "core" wealthy countries), an audience centric approach would likely reveal a more decentralized Web structure driven largely by culturally defined patterns of consumption. In other words, it is hypothesized that global online audience traffic would reflect culturally defined markets to a greater extent than hyperlink networks do.

Audience duplication is not the same as clickstreams (used by Wu and Ackland, 2014). For a website pair (i, j) the former includes all audiences who visited both these websites, whereas the latter counts audiences who immediately went to website i before or after website j. Presumably the duplication levels between website pairs are higher than their incoming or outgoing clickstreams, since the total number of people visiting any two websites would be greater than or equal to people visiting them in succession. Hence the audience duplication network would also be denser than the one based on clickstreams. Consequently, I expect even greater divergence between hyperlinks and audience network than Auckland and Wu (2014) found between hyperlinks and clickstreams.

## Method

I obtained audience duplication figures between all possible pairs of the most popular Websites. In doing so, duplication between any two Websites was operationalized as the percentage of unique users that visited both Websites across any possible pair. This measure reveals the extent to which any pair of Websites has audiences in common, irrespective of their language or the geography they focus on. I then used the resulting audience duplication matrix was then used to construct an "audience map" of the Web. I then compared this structure with a similar matrix obtained through hyperlink analysis. Next, I specify the data sources for both these matrices.

**Data**

**Audience network**. In this study, I use data from comScore[1], a panel based service that provides Internet audience measurement data once a month. It is currently the largest continuously measured audience panel of its kind. With approximately 2 million consumers worldwide in 170 countries under continuous measurement, the comScore panel utilizes a meter that captures behavioral information through a panelist's computer. Data are collected from both work and home computers of the panel members. Complementing the panel is a census-level data collection method, which allows for the integration of the aggregate level Internet behavior obtained through servers with audience information gained through the comScore panel.

ComScore organizes Websites by Web domains and subdomains. The sample for this study was chosen to be the top 1000 Web domains (ranked by monthly unique users) as this number not only captures most of the domains that 99% of Web users visit, but ensures an adequate representation of sites in different languages and different geographies. For many large Websites such as Google, the different geo-linguistic variants are classified as separate domains (e.g., www.google.es, www.google.de etc.) For certain large domains such as Wikipedia, language versions are sub-domains of the main domain (e.g., es.wikipedia.org). In such cases, these sub-domains have been considered in the final sample instead of restricting to top-level domains. These data reflect traffic during June 2012 (973 Websites of the top 1000 were included in the final sample.) These covered 50 languages in all (many sites were in multiple languages). For each one of the 973 Websites, I obtained its audience duplication with all other 972 sites. Thus the final dataset has 472878 ((973 *972)/2) pairs of audience duplication numbers.

**Hyperlinks network.** In addition to the traffic data, I obtained data on hyperlinks between Websites using a crawler called VOSON (Ackland, 2013). This crawler allows the user to specify all the Website addresses one needs hyperlinks on and downloads them systematically. Later it analyzes all the links present on the downloaded pages to provide hyperlinks between all pairs of initially specified Websites. Further it also queries a search engine, Blekko, which archives information on incoming links to Websites. Popular search engines such as Google and Bing no longer provide an API for this. This helps capture any links missed in the original downloads done by VOSON (as already noted, crawlers tend to miss out hyperlinks from social media sites). Owing to the sample having a large number of highly popular sites, a large number of pages had to be downloaded. Hence, the creators of VOSON provided the author with premium access and programming support to be able to perform an uninterrupted large-scale crawl that was done in August 2012. This resulted in a matrix that contained the number of hyperlinks between all possible pairs of 961 sites drawn from the same sample of 1000 most popular domains as the audience duplication. Some sites do not allow crawling and hence had to be dropped. The two matrices provided largely comparable data on hyperlinks and audience duplication between the same pair of most popular Websites.

**Language and geography.** All language(s) that each Website offered content in were noted by visiting each website. Further each Website's focal geographies were determined basis their languages, Wikipedia descriptions, information from their "About Us" pages, and site information on Alexa.com.

## Analysis and findings

I used the audience duplication matrix to conceptualize the Web as an "audience centric network" with the 973 Websites as nodes and % of unique users of total Internet audiences who visit both Websites as the ties. However since any two Websites are expected to have some amount of audience overlap, I considered a tie to exist only when the duplication was above what one would expect by random chance (see Taneja and Wu, 2014). For instance, if in a given time period, a certain Website, say 'A' has a unique user reach of 10% of all Internet users and Website 'B' has a reach of 50%, then assuming the consumption of both are independent events, 5% would be the expected number of users to visited both A and B. For all such pairs, a tie was considered only if the observed duplication was greater than the expected value. Hence the value of ties in the network is the *greater than expected duplication* between Websites. For some basic network measures, I used *dichotomous ties*, i.e., tie weights were assigned the value of '1' when present (i.e., when the observed duplication was greater than expected) and a value of '0' when absent.

Likewise, I symmetrized the matrix of hyperlinks between Websites to make it comparable to the audience matrix. This transformation converted the directed hyperlink network to an undirected network, but made it comparable to the audience network. In a directed network, each pair of nodes, say A and B, can have two possible connections, from A to B and B to A. No such directionality in a link is specified in an undirected network. Among online social networks, twitter ties are directed (a user can follow another user, but may not be followed by the same user in return). Facebook friendships, on the contrary, lead to an undirected network, as friendships on the service are reciprocal. For some analysis, I retained the tie values (i.e., number of hyperlinks in either direction), for others I *dichotomized* them (if any two Websites say A and B had at least one hyperlink between them in either direction they were considered as having a tie).

Before reporting the findings, I would like to explain some of the network analytic measures used. First *density* is the number of ties present as a proportion (or percentage) of the maximum number of ties possible. For instance, in a network of 10 nodes, a total of 45 undirected ties could be present (10 *9 /2) and if actually there were 30 ties, the network would have a density of 66% or .66. *Degree* of a node is the number of other nodes it has a tie with and a high degree score usually implies a more central node. *Clustering coefficient* indicates whether the network has a tendency to form communities of nodes such that nodes are tied with most other nodes within a community, but to relatively few nodes in other communities. It lies between '0' and '1' with 0 indicating no clustering and 1 being the score of a completely clustered network. Finally, *network centralization* represents the extent to which ties from most nodes are concentrated to a small number of nodes and can be considered analogous to Hirschman Herfindahl index, often used to measure market concentration (Webster and Ksiazek, 2012).

Comparing some descriptive statistics of the two networks (using dichotomized ties) on the measures just described confirmed the theoretical expectations. The hyperlink network has a low clustering coefficient (Table 1) and a degree distribution that reveals the presence of a small number of hubs (Figure 3). In other words, a few sites receive and send a lion's share of hyperlinks to most other sites. These are as expected the most popular Websites (e.g., Google.com, Facebook.com, Baidu.com) in terms of unique user numbers. The audience network on the other hand is a much denser network, with clear evidence of clustering, lower centralization and a relatively more uniform degree distribution (Table 1 and Figure 4). The measures reported so far corresponded to the dichotomized network. For the cluster and correlation analysis described as follows, I retained the values of tie weights in both networks.

To detect clusters, I ran a modularity-based community detection algorithm on both these networks, retaining the tie weights. The audience graph revealed 9 communities based on geo-linguistic similarities. Specifically, these clusters were of sites in English (focusing on US, Canada, UK, India and Australia), Chinese, French, Japanese & Korean, Portuguese (mainly Brazilian sites), Russian, German-Polish and Turkish. These findings certainly are consistent with the perspectives on global media flows that posit global audience flows to shape as culturally defined markets, as were evidenced in case of television and film audiences and have been more recently observed for the Web (Taneja and Wu, 2014; Taneja and Webster, 2016).

The hyperlinks network by contrast revealed 8 large communities (and several smaller clusters of dyads, traids etc). Except a cluster of Chinese websites, language and/or geography cannot explain website membership in these clusters. These appear more driven by either webmasters generally linking to prominent sites or cross- linking patterns between websites owned by the same publisher. For instance, one such community contained several language Wikipedias, whereas in the audience network each language Wikipedia clustered within their own language clusters. Likewise, country specific versions of sites such as Amazon, eBay, yahoo and Google, grouped together in clusters derived from the hyperlinks network, whereas in the audience network these clustered with their own language communities.

Figures 1 and 2 respectively represent the visualizations of the hyperlink and audience networks, using a class of layout techniques known as Fruchterman Reingold force directed algorithms. The colors in each of the graphs represent the geography (country) of the Website's focus, and the number used as the node label indicates cluster membership. In other words, websites having the same number clustered together, whereas websites having the same color focus on the same geography (country). Figure 2 affirms that clustering of the hyperlink graph did not reveal meaningful geo-linguistic clusters. As expected from prior work based on WST, hyperlinks exhibit a network structure with high centralization, whereas the audience network shows much more evidence of a decentralized structure driven by geo-linguistic clustering with moderate centralization and no clearly evident central core.

**Table 1 about here.**

**Figures 1 2, 3 and 4 about here**

Finally, as a confirmatory step, I obtained a measure of correlation between *valued* ties in the two networks. This measure provided the extent to which the number of hyperlinks between any two Websites correlated with the extent of (greater than expected) audience overlap between them. 950 sites common to either network were exactly comparable and were included for the analysis (23 sites had to be dropped due to incompatible domain architecture). A correlation analysis using a *Quadratic Assignment Procedure* (Krackardt, 1987) revealed a correlation coefficient (r) of 0.077 (*p<0.001*), suggesting that the number of hyperlinks between any pair of Websites is largely uncorrelated with the extent of audience duplication between them.

**Discussion**

This study compares the structure of the Web obtained through two approaches, the first being hyperlinks between Websites and the second an audience centric approach that relies on the extent to which two sites have audiences in common. The visualizations, descriptive statistics and cluster analysis of the two networks, and the (lack of) correlations between them, all suggest that the hyperlinks structure of the Web significantly differs from the usage patterns of online audiences. In other words, this analysis confirms the hypothesis that maps of Web based on actual usage do not mirror the structure of hyperlinks.

An analogy between airline traffic/routes and online audience traffic/hyperlinks is helpful in understanding this finding. The mere existence of a flight route between two airports does not indicate the volume of air traffic between them. A map of flight connections would reveal a few global hubs in the developed world that connect to regional airports in the periphery. However, more traffic circulates within a region or country than between regions or countries. Hence instead of seeing a few global hubs as evidenced when looking at routes alone, we see many more airports that serve as national/regional hubs central within their specific geographical clusters. I observe a similar discrepancy between the hyperlinks network (analogous to mapping routes) and the audience network (analogous to mapping airline traffic).

Before dwelling into the implications of these findings, it is worth speculating into the reasons for divergence between the hyperlinks and audience duplication networks. As already noted, Webmasters don't necessarily provide hyperlinks between Websites to direct traffic, but for a variety of reasons such as to signal affiliation, legitimacy, or merely refer to related content. The datasets analyzed here do show instances of such motivations. For instance, the hyperlink network analyzed had many links between CNN.com (the US domestic edition), CNN 's international edition and CNN's Spanish edition focused on Mexico. However, the audience network revealed little duplication between these sites. This does suggest that although CNN provides a lot of hyperlinks between its affiliate Websites, the same users do not access more than one of these sites. This is plausible as each of these Websites focuses on different regions of the world and recent studies show that Web use is geographically determined (Zuckerman, 2013). A similar divergence was seen in the case of Wikipedia, where editors provide hyperlinks between various language versions of the same concepts, but, as one would expect, on average, majority of users do not appear to visit Wikipedia in more than one language. Another reason for this observed divergence is that many hyperlinks become stale. This happens when webmasters do not remove links even after the target pages become inactive. Hyperlinks crawlers capture such stale links, even though audiences can no longer use these to navigate. Hyperlinks networks would include such links, but audience networks wouldn't.

Overall, the data showed very low correlation (7%) between the presence of pairs of hyperlinks, between pairs of Websites, and the presence of duplicated audiences between them. Perhaps language, geography and content similarities would each explain a large part of traffic. However, users also rely on many other structural features of the Web to navigate between Websites. These include search engines, recommendation systems, bookmarks, and a variety of such information regimes used by Websites (Webster, 2010). Future research should incorporate the role of these in explaining navigation patterns alongside hyperlinks. Adding to the discrepancy is the limitation with Web crawls, which as already noted, do not retrieve hyperlinks from semi-private Websites, such as Facebook and twitter. Further, users may at times access two websites in common, when a third site may hyperlink to both. This audience duplication based approach followed in this study relies on pair-wise comparisons is unable to decipher this detail.

Whatever the reasons, this divergence between hyperlink structures of and audience duplication patterns on the Web has important theoretical and practical implications. In the remainder of this section, I outline some of these.

First, as already noted, studies based on hyperlink analysis have painted somewhat dystopian perspectives of the Web's structure. These range from how Web replicates the core periphery structure of nation states as per the WST to how it reinforces imperialistic domination of media in the developing world by wealthy and large nations. An underlying assumption in this literature considers a country's position on the hyperlinks graph as a proxy for its potential for development, and predictors of their interactions with other countries (e.g., Park et al., 2011). However, the findings here suggest that such theorizations of cyberspace, based on hyperlink analysis alone, although "offer hybrid conceptualizations of the relationships between space and cyberspace"; they represent rather "simplistic understandings of the Internet and its effects [which] continue to inform a variety of economic development theories, especially those relating to e-commerce and commodity chains" (Graham, 2008, p. 775).

The structure of the Web obtained from audience duplication, on the contrary, exhibits the tendency of audiences to gravitate towards culturally proximate content. Hyperlink analysis although useful belongs to a category of many such maps of the Web infrastructure, which critical scholars have rightly warned masks aspects of the Web usage as it imagines the Web solely from the perspective of network centers (Burrell, 2012). The divergence between the simultaneously mapped hyperlinks and audience network in this study reiterates their stance. It further demonstrates that "geography has always been relational, and technology can therefore only ever supplement place-based existence instead of replacing it" (Graham, 2008: 774)

Second, the study demonstrates the use of audience duplication as an analytical tool in an alternate user-centric approach to mapping the Web. This conceptualization has the potential to explain mechanisms driving behavior of online audiences that cannot be revealed by hyperlink analysis, or for that matter any approach that focuses on infrastructure that imagines the Web solely as a globally connected, spaceless, borderless medium (Graham, 2013). Instead, mapping the Web as a network of shared audience traffic reveals the contours of how as divergent social groups join the Web; they take over its topography heterogeneously and unevenly, revealing aspects that cannot be fathomed from maps of technical features alone (Burrell, 2012).

Despite the stated virtues, these audience centric maps of the Web do have their limitations. The nodes in the audience network used in this study are at the Web domain level (comparable to the hyperlinks network). However, many Web domains labeled "global" are essentially centralized platforms with users in multiple countries across continents. Hence, with the current dataset one cannot for instance distinguish between Facebook users in US, Korea or Myanmar. However, being able to disaggregate such usage would- based on the findings in this study -reveal an even more decentralized audience-centric map of the Web.

In sum, the study contributes to emerging scholarship on global Web usage which has departed from focusing on infrastructure alone and provided a more audience centric conceptualization of how the Web is structured. These include studies on global networks of emails, (State et al., 2015) Twitter (Takhteyev et al., 2012), and clickstreams (Wu and Ackland, 2014); findings from these suggest that audience behavior on the net does not necessarily follow the medium's technical protocols, but language and geography have a prominent place in shaping global Web usage. The present study builds on and extends these studies to examine 99% of all Web traffic by considering all of the 1000 most visited domains irrespective of language, geography and content type. In sum, this study not only contributes to this growing body of literature on global Web usage, but also demonstrates the utility of using an audience duplication based approach alongside other approaches such as hyperlink analysis to explore the structure of the Web.

# Endnotes

[1] This information has been taken from ComScore's own documentation on methodology that is available through the help menu within the tool only to subscribers.

**Table 1: Descriptive network statistics**

| Statistic | Hyperlink Network (N=961) | Audience Network (N=973) |
|---|---|---|
| Clustering Coefficient | 0.524 | 0.846 |
| Network Centralization | 68% | 52% |
| Density* | 0.048 | 0.395 |

*Density of hyperlinks network would be even lower if it were not made symmetric*

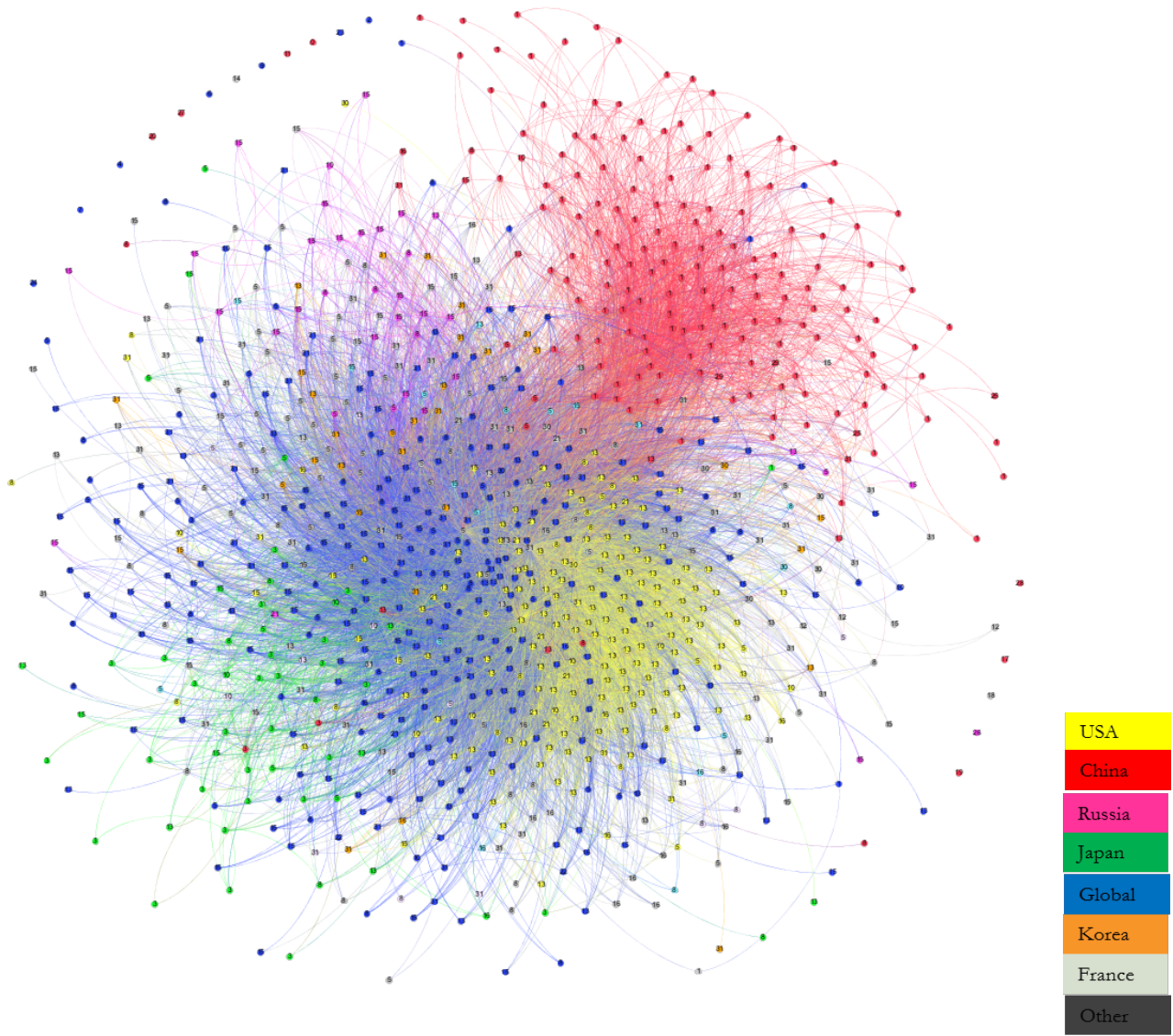

**Figure 1 Hyperlinks network**

*(In Figure 1 and Figure 2, "Global" refers to sites where a focal geography cannot be established. These are usually platforms in multiple languages and geographies)*

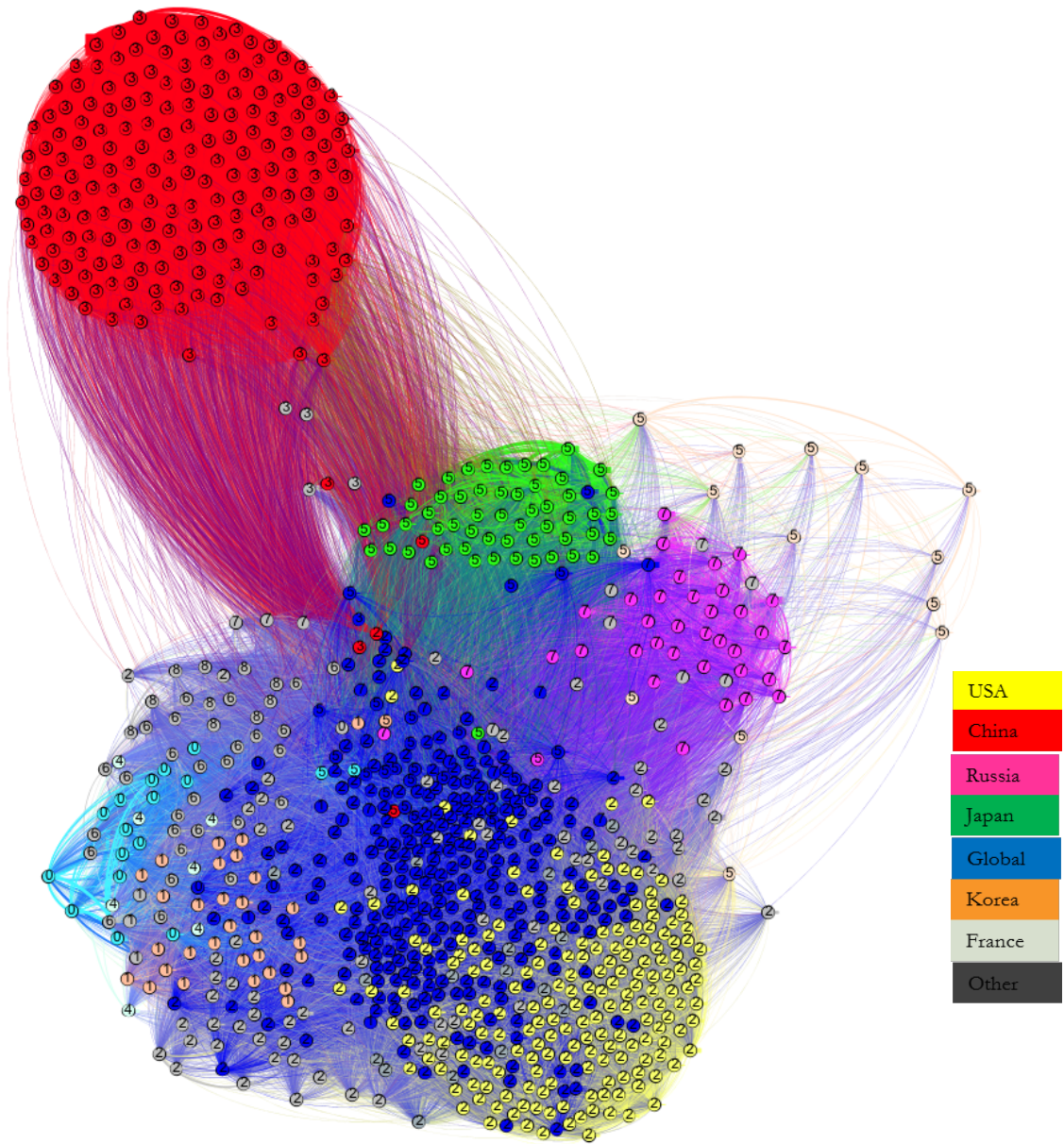

**Figure 2 Audience network**

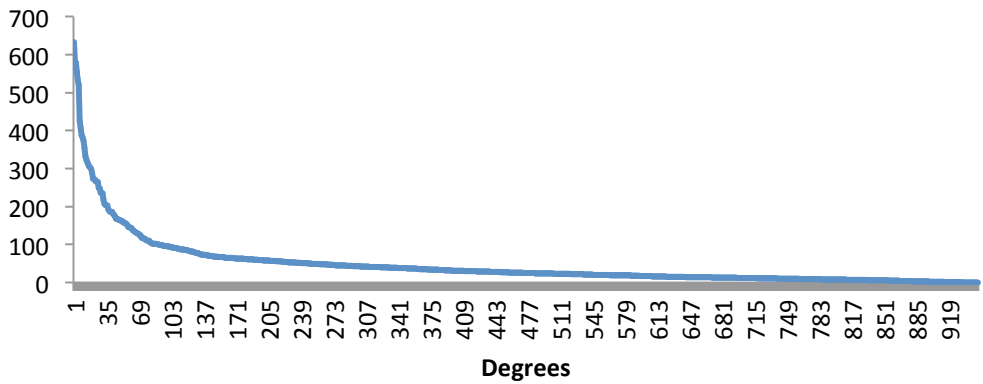

**Figure 3: Degree distribution of hyperlinks network**

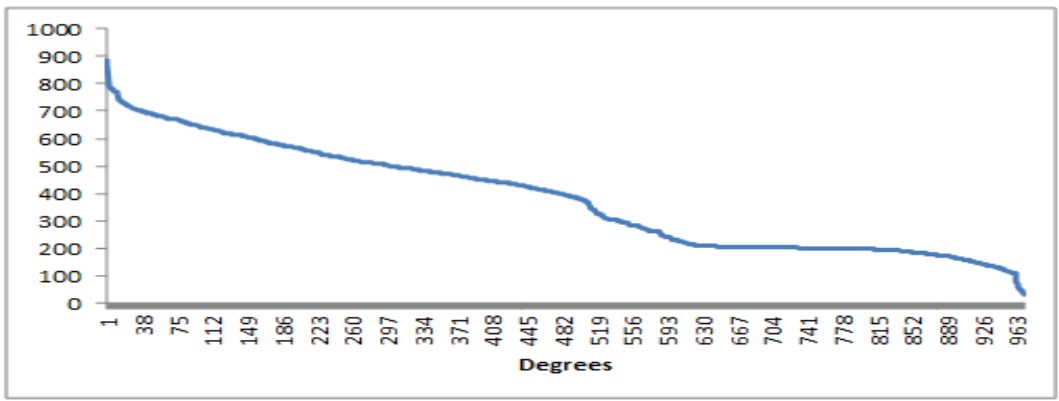

**Figure 4: Degree distribution of audience network**